\documentclass[aps,prl,twocolumn,superscriptaddress,showpacs,floatfix,nofootinbib]{revtex4-1}
\usepackage{bm,amsmath,graphicx}
\begin{document}

\title{Universal fluctuation-driven eccentricities in proton-proton, proton-nucleus and nucleus-nucleus collisions}

\author{Li Yan}
\email{li.yan@cea.fr}
\author{Jean-Yves Ollitrault}
\email{Jean-Yves.Ollitrault@cea.fr}
\affiliation{
CNRS, URA2306, IPhT, Institut de physique th\'eorique de Saclay, F-91191
Gif-sur-Yvette, France} 
\date{\today}

\begin{abstract}
We show that the statistics of fluctuation-driven initial-state
anisotropies in proton-proton, proton-nucleus and nucleus-nucleus collisions is to a
large extent universal. We propose a simple parameterization for 
the probability distribution of the Fourier coefficient 
$\varepsilon_n$ in harmonic $n$, which is in good agreement with 
Monte-Carlo simulations.
Our results provide a simple explanation for the 4-particle cumulant 
of triangular flow measured in Pb-Pb collisions, and for the 
 4-particle cumulant of elliptic flow recently measured in p-Pb collisions. 
Both arise as natural consequences of
the condition that initial anisotropies are bounded by unity. 
We argue that the initial rms anisotropy in harmonic $n$ can be
directly extracted from the measured ratio $v_n\{4\}/v_n\{2\}$: this gives
direct access to a property of the initial density profile from 
experimental data. 
We also make quantitative predictions for the small lifting of degeneracy 
between  $v_n\{4\}$, $v_n\{6\}$ and  $v_n\{8\}$.
If confirmed by future experiments, they will 
support the picture that long-range correlations observed
in p-Pb collisions at the LHC originate from collective flow
proportional to the initial anisotropy. 
\end{abstract}


\maketitle
\section{Introduction}
A breakthrough in our understanding of high-energy nuclear collisions
is the recognition~\cite{Alver:2006wh,Alver:2010gr}
that quantum fluctuations in the wavefunctions of
projectile and target, followed by hydrodynamic expansion, result in 
unique long-range azimuthal correlations between outgoing particles. 
The importance of these fluctuations was pointed out in the context of
detailed analyses of elliptic flow in nucleus-nucleus 
collisions~\cite{Alver:2006wh,Miller:2003kd}. 
It was later realized that fluctuations produce triangular
flow~\cite{Alver:2010gr}, which has subsequently been measured in nucleus-nucleus
collisions at RHIC~\cite{Adare:2011tg,Adamczyk:2013waa} and LHC~\cite{ALICE:2011ab,Chatrchyan:2012wg,ATLAS:2012at}. 
Recently, fluctuations were predicted to generate significant 
anisotropic flow in proton-nucleus
collisions~\cite{Bozek:2011if}, which quantitatively
explains~\cite{Bozek:2012gr}  the long-range correlations
observed by LHC experiments~\cite{CMS:2012qk,Abelev:2012ola,Aad:2012gla}. 

Recently, the ATLAS and CMS experiments reported the observation of a
nonzero  4-particle cumulant of azimuthal correlations, dubbed
$v_2\{4\}$, in proton-nucleus
collisions~\cite{Aad:2013fja,Chatrchyan:2013nka}.  
The occurrence of a large $v_2\{4\}$ in proton-nucleus collisions is
not fully understood, even though it is borne out by hydrodynamic
calculations with fluctuating initial conditions~\cite{Bozek:2013uha}.  
Such higher-order cumulants were originally 
introduced~\cite{Borghini:2000sa,Borghini:2001vi}
to measure elliptic flow in the reaction plane of non-central
nucleus-nucleus collisions, and isolate it from other, ``nonflow''
correlations. 
It turns out that the simplest fluctuations one can think of, namely, 
Gaussian fluctuations, do not contribute to 
$v_2\{4\}$~\cite{Voloshin:2007pc}.  
Since flow in proton-nucleus collisions is thought to originate from 
fluctuations in the initial geometry, 
one naively expects $v_2\{4\}\sim 0$, {\it even if\/}
there is collective flow in the system. 

In this paper, we argue that the values observed for $v_2\{4\}$ in p-Pb collisions 
are naturally explained by non-Gaussian fluctuations, which are expected for small systems. 
Our explanation differs from that recently put forward by 
Bzdak et al.~\cite{Bzdak:2013rya} that it is due to symmetry breaking
(see Eq.~(\ref{bg}) and discussion below). 
As Bzdak et al., we assume that anisotropic flow $v_n$ scales like the
corresponding initial-state anisotropy $\varepsilon_n$ on an
event-by-event basis. This is known to be a very good approximation in
ideal~\cite{Qiu:2011iv} and viscous~\cite{Niemi:2012aj} hydrodynamics. 
Thus flow fluctuations directly reflect $\varepsilon_n$ fluctuations. 
Now, $\varepsilon_n$ is bounded by unity by definition. 
On the other hand, Gaussian fluctuations are not bounded, which is the
reason why they fail to model small systems. 
We propose a simple alternative to the Gaussian parameterization 
which naturally satisfies the constraint $\varepsilon_n<1$. We show
that it provides an excellent fit to all Monte-Carlo 
calculations.

\section{Distribution of the initial anisotropy}
In each event, the anisotropy in harmonic $n$ is defined (for $n=2,3$) by~\cite{Teaney:2010vd}
\begin{eqnarray}
\label{defepsilon}
\varepsilon_{n,x} &\equiv& -\frac{\int r^n \cos(n\phi)\rho(r,\phi)rdrd\phi}
{\int r^n\rho(r,\phi)rdrd\phi}\cr
\varepsilon_{n,y} &\equiv& -\frac{\int r^n \sin(n\phi)\rho(r,\phi)rdrd\phi}
{\int r^n\rho(r,\phi)rdrd\phi},
\end{eqnarray}
where $\rho(r,\phi)$ is the initial transverse density profile near midrapidity 
 in a centered polar coordinate system.
 
\begin{figure}
 \includegraphics[width=\linewidth]{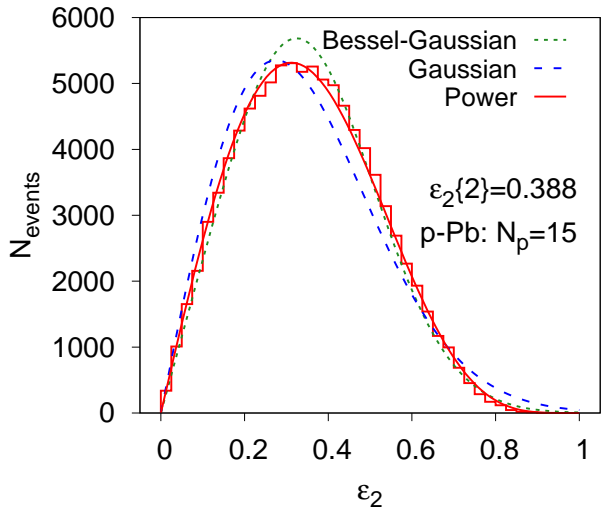}
 \caption{ (Color online) 
Histogram of the distribution of 
$\varepsilon_2$ obtained in a Monte-Carlo Glauber simulation of a p-Pb
collision at LHC, and fits using 
Eqs.~(\ref{gaussian})-(\ref{powerlaw}). }  
\label{fig:histogram}
\end{figure}

Fig.~\ref{fig:histogram} displays the histogram of the distribution of
$\varepsilon_2$ in a p-Pb collision at 5.02~TeV obtained in a Monte-Carlo Glauber
calculation~\cite{Miller:2007ri}. 
We use the PHOBOS implementation~\cite{Alver:2008aq} with a Gaussian
wounding profile~\cite{Alvioli:2009ab,Rybczynski:2011wv}. 
We assume that the initial density $\rho(r,\phi)$ is a sum of
Gaussians of width $\sigma_0=0.4$~fm, centered around each participant
nucleon with a normalization that
fluctuates~\cite{Rybczynski:2013yba}.  
These fluctuations, which increase anisotropies~\cite{Dumitru:2012yr}, 
are modeled as in Ref.~\cite{Bzdak:2013rya}. 
We have selected events with number of participants $14\le N\le 16$, 
corresponding to typical values in a central p-Pb collision. 

We now compare different parameterizations of this distribution, which
we use to fit our numerical results. 
The first is an isotropic two-dimensional Gaussian (we drop the
subscript $n$ for simplicity):
\begin{equation}
\label{gaussian}
P(\varepsilon)=\frac{2\varepsilon}{\sigma^2}\exp\left(-\frac{\varepsilon^2}{\sigma^2}\right),
\end{equation}
where $\varepsilon\equiv\sqrt{\varepsilon_x^2+\varepsilon_y^2}$ and
the distribution is normalized:
$\int_0^{\infty}P(\varepsilon)d\varepsilon=1$.
This form is motivated by the central limit theorem, 
assuming that the eccentricity solely originates from event-by-event
fluctuations, and neglecting fluctuations in the denominator. 
Note that this distribution does not strictly satisfy  the constraint
$\varepsilon<1$, which follows from the definition
(\ref{defepsilon}). 
When fitting our Monte-Carlo results, we have therefore multiplied
Eq.~(\ref{gaussian}) 
by a constant to ensure normalization between 0 and 1. 
The rms $\varepsilon$ has been fitted to that of the Monte-Carlo
simulation. 
Fig.~\ref{fig:histogram} shows that Eq.~(\ref{gaussian}) gives a
reasonable approximation to our Monte-Carlo results, but not a good
fit. 

Bzdak et al.~\cite{Bzdak:2013rya} have proposed to replace Eq.~(\ref{gaussian})
by a ``Bessel-Gaussian'':
\begin{equation}
\label{bg}
P(\varepsilon)=\frac{2\varepsilon}{\sigma^2}
I_0\left(\frac{2\varepsilon\bar\varepsilon}{\sigma^2}\right)
\exp\left(-\frac{\varepsilon^2+\bar\varepsilon^2}{\sigma^2}\right).
\end{equation}
This parameterization introduces an additional free parameter $\bar\varepsilon$, 
corresponding to the mean eccentricity in the reaction plane in nucleus-nucleus 
collisions~\cite{Voloshin:2007pc}. 
It reduces to (\ref{gaussian}) if  $\bar\varepsilon=0$. 
A nonzero value of $\bar\varepsilon$ is however difficult to justify for a 
symmetric system in which anisotropies are solely created by fluctuations. 
In Fig.~\ref{fig:histogram}, $\bar\varepsilon$ and $\sigma$ have been
chosen so that the first even moments $\langle\varepsilon^2\rangle$
and $\langle\varepsilon^4\rangle$ match exactly the Monte-Carlo
results, as suggested in~\cite{Bzdak:2013rya}. 
The quality of the fit is not much improved compared to the Gaussian
distribution, even though there is an additional free parameter. 
Note that the Bessel-Gaussian, like the
Gaussian, does not take into account the constraint $\varepsilon<1$.

We now introduce the one-parameter power law distribution: 
\begin{equation}
\label{powerlaw}
P(\varepsilon)=2\alpha\varepsilon(1-\varepsilon^2)^{\alpha-1},
\end{equation}
where $\alpha>0$. 
Eq.~(\ref{powerlaw}) reduces to Eq.~(\ref{gaussian}) for 
$\alpha\gg 1$, with $\sigma^2\equiv 1/\alpha$. 
The main advantage of Eq.~(\ref{powerlaw}) over previous
parameterizations is that the support of  $P(\varepsilon)$ is the
unit disc: it satisfies for all $\alpha>0$ the 
normalization $\int_0^1 P(\varepsilon)d\varepsilon=1$. 
In the limit $\alpha\to 0^+$,   $P(\varepsilon)\simeq\delta(\varepsilon-1)$. 

Eq.~(\ref{powerlaw}) is the {\it
  exact\/}~\cite{Ollitrault:1992bk}\footnote{See Eq.~(3.10) 
  of~\cite{Ollitrault:1992bk}. What is derived there is the
  distribution of anisotropy in momentum space, but the algebra is
  identical for the distribution of eccentricity.} distribution of 
$\varepsilon_2$ for $N$ identical pointlike sources with a
2-dimensional isotropic Gaussian  distribution, with $\alpha=(N-1)/2$, 
if one ignores the recentering correction. 
In a more realistic situation, Eq.~(\ref{powerlaw}) is no longer
exact. We adjust $\alpha$ to match the rms $\varepsilon$ from the
Monte-Carlo calculation. 
Fig.~\ref{fig:histogram} shows that Eq.~(\ref{powerlaw}) (with
$\alpha\simeq 5.64$) agrees much
better with Monte-Carlo results than Gaussian and
Bessel-Gaussian distributions. 

\section{Cumulants}
Cumulants of the distribution of $\varepsilon$ are derived from a
generating function, which is the logarithm of the two-dimensional
Fourier transform of the distribution of $(\varepsilon_x,\varepsilon_y)$: 
\begin{equation}
\label{defg}
G(k_x,k_y)\equiv \ln \langle \exp(ik_x \varepsilon_x+ik_y\varepsilon_y)\rangle,
\end{equation}
where angular brackets denote an expectation value over the ensemble of events. 
If the system has azimuthal symmetry, by integrating over the relative
azimuthal angle of ${\bf k}$ and $\boldsymbol{\varepsilon}$, one obtains 
\begin{equation}
\label{defgbessel}
G(k)=\ln \langle J_0(k\varepsilon)\rangle,
\end{equation}
where $k\equiv\sqrt{k_x^2+k_y^2}$ and
$\varepsilon\equiv\sqrt{\varepsilon_x^2+\varepsilon_y^2}$. 
The cumulant to a given order $n$, $\varepsilon\{n\}$, is obtained by
expanding Eq.~(\ref{defgbessel}) to order $k^n$,
and identifying with the expansion of $\ln J_0(k\varepsilon\{n\})$ to
the same order. 
This uniquely defines $\varepsilon\{n\}$ for all {\it even\/} $n$. 
One thus obtains~\cite{Miller:2003kd}
$\varepsilon\{2\}^2=\langle\varepsilon^2\rangle$, 
$\varepsilon\{4\}^4=2\langle\varepsilon^2\rangle^2-\langle\varepsilon^4\rangle$. 
Expressions of $\varepsilon\{6\}$ and $\varepsilon\{8\}$ are given
in~\cite{Bzdak:2013rya}.

\begin{table}[ht]
\caption{Values of the first eccentricity cumulants 
for the Gaussian (\ref{gaussian}), Bessel-Gaussian (\ref{bg}) and
power law 
 (\ref{powerlaw}) distributions. }
\label{table:models}
\smallskip
\begin{center}
\begin{tabular}{|l|c|c|c|}
\hline
&Gauss&BG&Power\\
\hline
$\varepsilon\{2\}$&$\sigma$&$\sqrt{\sigma^2+\bar\varepsilon^2}$&$\frac{\displaystyle
  1}{\displaystyle\sqrt{1+\alpha}}$\\
\hline
$\varepsilon\{4\}$&0&$\bar\varepsilon$&
$\left[\frac{\displaystyle 2}{\displaystyle (1+\alpha)^2(2+\alpha)}\right]^{1/4}$\\
\hline
$\varepsilon\{6\}$&0&$\bar\varepsilon$&
$\left[\frac{\displaystyle 6}{\displaystyle (1+\alpha)^3(2+\alpha)(3+\alpha)}\right]^{1/6}$\\
\hline
$\varepsilon\{8\}$&0&$\bar\varepsilon$&
$\left[\frac{\displaystyle 48\left(1+\frac{5\alpha}{11}\right)}
{\displaystyle (1+\alpha)^4(2+\alpha)^2(3+\alpha)(4+\alpha)}\right]^{1/8}$\\
\hline
\end{tabular}
\end{center}
\end{table}
Expressions of the first four cumulants are listed in
Table~\ref{table:models}.
For the power law distribution (\ref{powerlaw}), these results are 
obtained by expanding the generating function (\ref{defgbessel}): 
\begin{eqnarray}
\label{gpowerlaw}
G(k)&=&\ln\left[\int_0^1
J_0(k\varepsilon)P(\varepsilon)d\varepsilon\right]\cr
&=&\ln\left[\frac{2^{\alpha}\alpha!}{k^{\alpha}}J_{\alpha}(k)\right]. 
\end{eqnarray}
General results have been obtained previously 
in the case of $N$ pointlike sources and in the large $N$ limit for 
$\varepsilon_2\{2\}$~\cite{Bhalerao:2006tp} and 
$\varepsilon_2\{4\}$~\cite{Alver:2008zza}. 
Our results derived from Eq.~(\ref{powerlaw}) are exact for a Gaussian distribution
of sources and therefore agree with these general results for $N\gg 1$. 
Similar results have also been derived for $\varepsilon_3\{2\}$ and
$\varepsilon_3\{4\}$~\cite{Bhalerao:2011bp}, but not for 
cumulants of order 6 or higher. 

\begin{figure}
 \includegraphics[width=\linewidth]{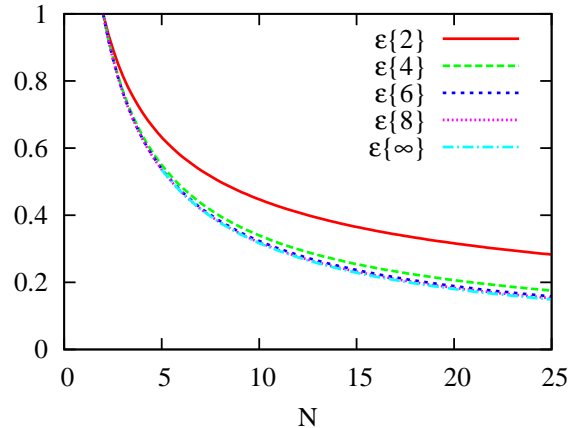}
 \caption{ (Color online) Cumulants of the eccentricity distribution
   as a function of the number of participants $N$ 
   for the power law distribution (\ref{powerlaw}), where we have 
set   $\alpha=(N-2)/2$. 
}  
\label{fig:gaussian}
\end{figure}

Fig.~\ref{fig:gaussian} displays the cumulants $\varepsilon\{2\}$ to
$\varepsilon\{8\}$ as a function of $N$, as
predicted by Eq.~(\ref{powerlaw}) for pointlike
sources.\footnote{Here, we
  assume that the recentering correction 
  effectively reduces   by one unit the number of independent
  sources. 
  We thus replace $N$ by $N-1$ in the exact result of 
  Ref.~\cite{Ollitrault:1992bk}.}
These results are similar to those obtained in full Monte-Carlo Glauber
calculations~\cite{Bzdak:2013rya}. 
In the limit  $N\gg 1$, the power law distribution yields 
$\varepsilon\{k\}\propto N^{(1-k)/k}$.
It thus predicts 
a strong ordering $\varepsilon\{8\}\ll \varepsilon\{6\}\ll \varepsilon\{4\}\ll \varepsilon\{2\}\ll 1$,
unlike the Bessel-Gaussian which predicts
$\varepsilon\{4\}=\varepsilon\{6\}=\varepsilon\{8\}$.
For fixed $N$, however, the cumulant expansion
quickly converges, as illustrated in Fig.~\ref{fig:gaussian}. 
In practice, for typical values of $N$ in p-Pb collisions, one
observes $\varepsilon\{4\}\simeq \varepsilon\{6\}\simeq \varepsilon\{8\}$, in
agreement with numerical findings of Bzdak et
al.~\cite{Bzdak:2013rya}. 
This rapid convergence can be traced back to the fact that the
generating function $G(k)$ in Eq.~(\ref{gpowerlaw}) has a singularity  
at the first zero of $J_\alpha(k)$, denoted by $j_{\alpha 1}$. 
This causes the cumulant expansion to quickly converge to the
value~\cite{Bhalerao:2003xf} 
\begin{equation}
\label{lyz}
\varepsilon\{\infty\}=\frac{j_{01}}{j_{\alpha 1}}. 
\end{equation}
This asymptotic limit is also plotted in Fig.~\ref{fig:gaussian}. It
is hardly distinguishable from $\varepsilon\{6\}$ and $\varepsilon\{8\}$ 
for these values of $N$. 

\begin{figure}
 \includegraphics[width=\linewidth]{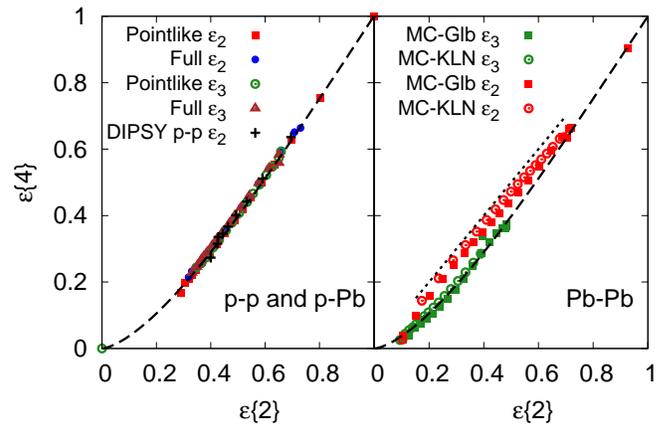}
 \caption{ (Color online) $\varepsilon\{4\}$ versus 
   $\varepsilon\{2\}$. The dashed line in both panels is
   Eq.~(\ref{eps4vseps2}). Left: p-Pb collisions. ``Full'' refers to
   Gaussian sources associated with each participant, and
   fluctuations in the weights of each source. ``Pointlike'' refers to
   pointlike identical sources. DIPSY results for p-p collisions are replotted
   from~\cite{Avsar:2010rf}. Right: Pb-Pb collisions. The dotted
   line is $\varepsilon\{4\}=\varepsilon\{2\}$, corresponding to a
   nonzero mean eccentricity, and negligible fluctuations.}  
\label{fig:e2e4}
\end{figure}

\section{Testing universality}
The power law distribution (\ref{powerlaw}) predicts the following
parameter-free relation between the first two cumulants:
\begin{equation}
\label{eps4vseps2}
\varepsilon\{4\}=\varepsilon\{2\}^{3/2}\left(\frac{2}{1+\varepsilon\{2\}^2}\right)^{1/4}.
\end{equation}
This relation can be used to test the universality of the
distribution (\ref{powerlaw}). 
For p-Pb collisions at 5.02~TeV, we run two different types of
Monte-Carlo Glauber 
calculations: a full Monte-Carlo identical to that of 
Fig.~\ref{fig:histogram}, and a second one where fluctuations and
smearing are switched off (identical pointlike sources). 
We calculate $\varepsilon_2$ and $\varepsilon_3$ for each event. 
Events are then binned according to the number of participants $N$,
mimicking a centrality selection. 
For p-p collisions at 7~TeV, we use published 
results~\cite{Avsar:2010rf} obtained with the event 
generator DIPSY~\cite{Flensburg:2011wx}, which are binned according 
to multiplicity. 
Results are shown in Fig.~\ref{fig:e2e4} (left). 
Each symbol of a given type corresponds to a different bin. 
All Monte-Carlo results are in very good agreement with Eq.~(\ref{eps4vseps2}). 
A closer look at the results show that the ``full'' Monte-Carlo 
Glauber calculations are above the line by $\sim 0.015$ (for both
$\varepsilon_2$ and $\varepsilon_3$), the 
``pointlike'' results for $\varepsilon_3$ by $\sim 0.005$, and the
``pointlike'' results for $\varepsilon_2$ (where our result is exact, up to the 
recentering correction) by $\sim 0.002$. 
DIPSY results are above the line by $\sim 0.01$.

For Pb-Pb collisions at 2.76~TeV (Fig.~\ref{fig:e2e4} right), we use the
results obtained in Ref.~\cite{Bhalerao:2011yg} using the Monte-Carlo
Glauber~\cite{Alver:2008aq} and Monte-Carlo KLN~\cite{Drescher:2007ax}
models. These results are in 5\% centrality bins.  
For $\varepsilon_3$, both models are in very good agreement with
Eq.~(\ref{eps4vseps2}) (within $0.01$ or so). Note that Pb-Pb
collisions probe this relation closer to the origin, in the large $N$
limit where more general results are
available~\cite{Bhalerao:2011bp}. These general  results predict
$\varepsilon\{4\}\propto\varepsilon\{2\}^{3/2}$ for $N\to\infty$, but
with a proportionality constant that depends on the density
profile. Our results show that it is in practice very close to 
the value predicted by Eq.~(\ref{eps4vseps2}), namely, $2^{1/4}$. 

Monte-Carlo results for $\varepsilon_2$ in Pb-Pb differ from
Eq.~(\ref{eps4vseps2}). This 
is expected, since $\varepsilon_2$ in mid-central Pb-Pb collisions is
mostly driven by the almond shape of the overlap area between colliding
nuclei~\cite{Ollitrault:1992bk}, not by fluctuations. 
In the limiting case where fluctuations are negligible,
$\varepsilon_2\{4\}=\varepsilon_2\{2\}$.
Our results show that fluctuations dominate only for the most central
and most peripheral bins. 

We conclude that the power law distribution (\ref{powerlaw}) is 
a very good approximation to the distribution of fluctuation-driven
eccentricities, irrespective of the details of the model. 
This could be checked explicitly with other initial-state 
models~\cite{Dumitru:2012yr,Schenke:2012hg}. 

\begin{figure}
 \includegraphics[width=\linewidth]{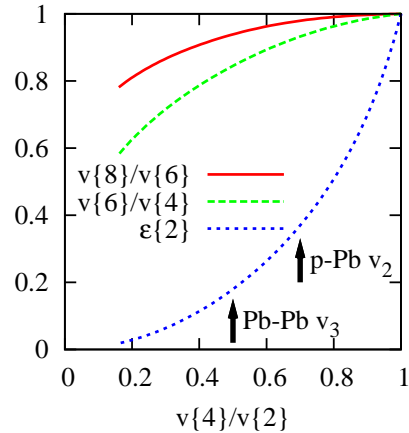}
 \caption{ (Color online) Predictions of the model for ratios of
   higher order cumulants and $\varepsilon\{2\}$ as a function of the measured
   $v\{4\}/v\{2\}$. Typical values for $v_3$ in
   Pb-Pb~\cite{ALICE:2011ab,Chatrchyan:2013kba}  and $v_2$ in p-Pb
   collisions~\cite{Chatrchyan:2013nka} are indicated by arrows. }  
\label{fig:predictions}
\end{figure}
\section{Applications} 
We now discuss applications of our result. 
The distribution of $\varepsilon_n$ is completely
determined by the parameter $\alpha$ in Eq.~(\ref{powerlaw}). This
parameter can be obtained directly from experimental data. 
Assuming that anisotropic flow is proportional to eccentricity in the
corresponding harmonic, $v_n\propto\varepsilon_n$, which is proven to
be a very good approximation for $n=2,3$~\cite{Niemi:2012aj}, one
obtains
\begin{equation}
\frac{v\{4\}}{v\{2\}}=\frac{\varepsilon\{4\}}{\varepsilon\{2\}}=
\left(\frac{2}{2+\alpha}\right)^{1/4}.
\end{equation}
The first equality has already been checked against Monte-Carlo models
and experimental data~\cite{Bhalerao:2011ry,Chatrchyan:2013kba}. The
second equality directly relates the parameter $\alpha$ in
Eq.~(\ref{powerlaw}) to the measured ratio $v\{4\}/v\{2\}$. 

This in turn gives a prediction for ratios of higher-order flow
cumulants, which scale like the corresponding ratios of eccentricity
cumulants. These predictions are displayed in
Fig.~\ref{fig:predictions}.  
One can also directly obtain the rms eccentricity $\varepsilon\{2\}$,
which is a property of the initial state. 

The ratio  $v_3\{4\}/v_3\{2\}$ in Pb-Pb is close to 
0.5 in mid-central collisions~\cite{ALICE:2011ab,Chatrchyan:2013kba}. 
We thus predict $v_3\{6\}/v_3\{4\}\simeq 0.84$ and
$v_3\{8\}/v_3\{6\}\simeq 0.94$ in the same centrality. 
We also obtain $\varepsilon_3\{2\}\simeq 
0.17$, which is a typical prediction from Monte-Carlo models 
in the 10\%-20\% or 20\%-30\% centrality
range~\cite{Retinskaya:2013gca}. 

Similarly, the ratio $v_2\{4\}/v_2\{2\}\sim 0.7$ measured in p-Pb
collisions~\cite{Aad:2013fja,Chatrchyan:2013nka} implies
$v_2\{6\}/v_2\{4\}\simeq 0.93$ and
$v_2\{8\}/v_2\{6\}\simeq 0.98$, that is, almost degenerate
higher-order cumulants. 
We obtain $\varepsilon_2\{2\}\simeq 0.37$, in agreement with 
Monte-Carlo Glauber models~\cite{Bzdak:2013rya}. 

\section{Conclusions} 
We have proposed a new parameterization of the distribution
of the initial anisotropy $\varepsilon_n$ in proton-proton,
proton-nucleus and nucleus-nucleus 
collisions which, unlike previous parameterizations, takes into
account the condition $\varepsilon_n<1$. 
This new parameterization is found in good agreement with results 
of Monte-Carlo simulations when 
$\varepsilon_n$ is created by fluctuations of the initial geometry. 
Our results explain the observation, in these Monte-Carlo models, that
cumulants of the distribution of $\varepsilon_n$ quickly converge as
the order increases. This is because the Fourier transform of the
distribution of $\varepsilon_n$ has a zero at 
a finite value of the conjugate variable $k$. This, in turn, is a
consequence of the fact that the probability distribution of
$\varepsilon_n$ has compact support (that is, $\varepsilon_n<1$). 

The consequence of this universality is that while the rms
$\varepsilon_n$ is strongly model-dependent~\cite{Retinskaya:2013gca}, 
the probability distribution of $\varepsilon_n$ is fully determined
once the rms value is known --- in particular, the magnitudes of
higher-order cumulants such as $\varepsilon_n\{4\}$. 
Assuming that anisotropic flow $v_n$ is proportional to
$\varepsilon_n$ in every event, we have predicted the values of
$v_3\{6\}$ and $v_3\{8\}$ in Pb-Pb collisions, and the values of
$v_2\{6\}$ and $v_2\{8\}$ in p-Pb collisions. 

If future experimental data confirm our prediction, these results will
strongly support the picture that the long-range correlations observed
in proton-nucleus and nucleus-nucleus collisions are due to
anisotropic flow, which is itself proportional to the anisotropy in
the initial state. This picture, furthermore, will be confirmed
irrespective of the details of the initial-state model.  

\begin{acknowledgments}
JYO thanks Art Poskanzer for pointing out, back in 2009, that Bessel-Gaussian fits 
to Monte-Carlo Glauber calculations fail because 
they miss the constraint $\varepsilon_2<1$, 
Larry McLerran for discussing Ref.~\cite{Bzdak:2013rya} prior to
publication, Christoffer Flensburg for sending DIPSY results, 
Ante Bilandzic and Wojciech Broniowski for useful discussions, 
and Jean-Paul Blaizot and Raju Venugopalan for comments on the manuscript. 
We thank the Yukawa Institute for Theoretical Physics, Kyoto University.
Discussions during the YITP workshop YITP-T-13-05 on ``New Frontiers in QCD'' were useful to complete this work.
LY is funded  by the European Research Council under the 
Advanced Investigator Grant ERC-AD-267258.
\end{acknowledgments}

\end{document}